\documentclass{aa}

\usepackage{graphicx}
\usepackage{txfonts}
\usepackage[normalem]{ulem}
\usepackage{natbib}

\usepackage{color}

\begin{document}

\title{A study of  the H{\,\sc i} gas fractions of galaxies at $z\sim1$} 
\titlerunning{H{\,\sc i} gas at high-z}

\author{Wei Zhang \inst{1},
        Guinevere Kauffmann \inst{2},
        Jing Wang \inst{3},             
        Yanmei Chen \inst{4,5},         
        Jian Fu \inst{6,7},             
        Hong Wu \inst{1}
       }
\authorrunning{Wei Zhang et al.}
\institute{CAS Key Laboratory of Optical Astronomy, National Astronomical Observatories, Chinese Academy of Sciences, Beijing 100101, China \\
           \email{xtwfn@bao.ac.cn}
           \and
           Max Planck Institut f\"ur Astrophysik, Karl-Schwarzschild-Strasse  1, 85748  Garching, Germany
           \and
           Kavli Institute for Astronomy and Astrophysics, Peking University, Beijing 100871, China
           \and
           School of Astronomy and Space Science, Nanjing University, Nanjing 210093, China
           \and
           Key Laboratory of Modern Astronomy and Astrophysics (Nanjing University), Ministry of Education, Nanjing 210093, China
           \and
           Shanghai Astronomical Observatory,  80 Nandan Road, Shanghai 200030, China
           \and
           Key Laboratory for Research in Galaxies and Cosmology, Shanghai Astronomical Observatory, CAS, 80 Nandan Road, Shanghai 200030, China
          }

\date{Received ........; accepted ........}

\abstract
{}
{Due to the fact that H{\,\sc i} mass measurements are not available for large
galaxy samples at high redshifts, we apply a photometric estimator of the H{\,\sc
i}-to-stellar mass ratio ($M_{\rm HI}/M_\ast$) calibrated using a local Universe
sample of galaxies to a sample of galaxies at $z\sim1$ in the DEEP2 survey.  We
use these H{\,\sc i} mass estimates to calculate H{\,\sc i} mass functions
(HIMFs) and cosmic H{\,\sc i} mass densities ($\Omega_{\rm HI}$), and to examine
the correlation between star formation rate and H{\,\sc i} gas content, for
galaxies at $z\sim1$.}
{We have estimated H{\,\sc i} gas masses for $\sim7,000$ galaxies in the DEEP2
survey with redshifts in the range $0.75<z<1.4$ and stellar masses
$M_\ast\gtrsim 10^{10}$\,$\rm M_\odot$, using a combination of the rest-frame
ultraviolet-optical colour $(NUV-r)$ and stellar mass density ($\mu_*$) as a
way to estimate $M_{\rm HI}/M_\ast$.}
{It is found that the high mass end of high-z H{\,\sc i} mass function (HIMF)
is quite similar to that of the local HIMF. The lower limit of $\Omega_{\rm HI,
limit}=2.1\times10^{-4}\, h_{70}^{-1}$, obtained by directly integrating the
H{\,\sc i} mass of galaxies with $M_\ast \gtrsim 10^{10}$\,$\rm M_\odot$, confirms
that massive star-forming galaxies do not dominate the neutral gas at $z\sim1$.
We study the evolution of the H{\,\sc i} mass to stellar
mass ratio from $z\sim1$ to today and find a steeper relation between
H{\,\sc i} gas mass fraction and stellar mass at higher redshifts.
Specifically, galaxies with $M_\ast = 10^{11}$\,$\rm M_\odot$ at z$\sim$1 are found to
have 3--4 times higher neutral gas fractions than local galaxies, while the
increase is as high as 4--12 times at $M_\ast = 10^{10}$\,$\rm M_\odot$.
The quantity $M_{\rm HI} /{\rm SFR}$ exhibits very large scatter,
and the scatter increases from a factor of 5--7 at $\rm z=0$ to factors 
close to a hundred at $\rm z=1$. This implies that there is no relation between
 H{\,\sc i} gas and star formation in high redshift galaxies. The H{\,\sc i} gas must be
linked to cosmological gas accretion processes at high redshifts.}
{}
 \maketitle

\section{introduction}
\label{sec:introduction}

Cold gas plays a crucial role in galaxy formation and evolution, yet our
understanding of the cold gas content of galaxies has been hampered for many
years by the lack of direct H{\,\sc i} line measurements for galaxies at high
redshifts. In the local Universe, much data on local galaxies exists from
surveys such as  the H{\,\sc i} Parkes All Sky Survey
\citep[HIPASS;][]{Meyer-04} and the Arecibo Legacy Fast ALFA (ALFALFA) survey
\citep{Giovanelli-05} of H{\,\sc i} 21-cm emission of galaxies. It should be
noted that these surveys are limited to nearby galaxies ($z\la 0.06$), and are
relatively shallow (H{\,\sc i}-to-stellar mass ratio $M_{\rm
HI}/M_\ast\ga10\%$). The extended GALEX Arecibo SDSS Survey
\citep[xGASS;][]{Catinella-18} was designed to observed galaxies to a fixed
detection limit in the quantity $M_{\rm HI}/M_\ast$ in order to detect small
amounts of residual gas in galaxies in transition between the star-forming
main sequence \citep{Noeske-07b} (star-forming galaxies that define the
relation between the star formation rate and stellar mass at a given redshift)
and the population of massive early-type galaxies at the same redshift where
star formation has largely ceased. This survey pushed the $M_{\rm HI}/M_\ast$
detection limit down to $\sim 1.5\%$ for a sample of 1179 galaxies with stellar
masses $10^9$\,$\rm M_\odot$ $<M_\ast<10^{11.5}$\,$\rm M_\odot$ and redshifts
in the range $0.01<z<0.05$.  For redshifts beyond this range, observations of
H{\,\sc i} in emission are available only for a handful of galaxies out to
$z<0.25$ \citep{Catinella-08}.  The current highest redshift of an individual
galaxy with a direct H{\,\sc i} emission detection at z = 0.376 is reported by
\cite{Fernandez-16} from a study with the COSMOS H{\,\sc i} Large Extragalactic
Survey.

At higher redshifts, observations of H{\,\sc i} in emission from
individual galaxies are lacking. There have been recent attempts to estimate
the {\em average} atomic gas content in star-forming galaxies at $z\sim0.24$
\citep{Lah-07} and $z\sim1.3$ \citep{Kanekar-16} and in cluster galaxies at
$z\sim0.37$ \citep{Lah-09}, by co-adding the 21-cm emission signals of few
hundred galaxies. \cite{Chowdhury-20} have applied the same method to a larger
sample of 7,653 star-forming galaxies at $z=0.74-1.45$. \citet{Pen-09}
proposed that the cosmic structure traced by atomic gas could be probed by
measuring the three-dimensional intensity map of the H{\,\sc i} gas in the
universe.  In a series of test-case applications, Pen and collaborators
measured  the clustering signal of H{\,\sc i} gas by cross-correlating H{\,\sc
i} data cubes with optical surveys of galaxies in both the local universe
\citep{Pen-09} and at $z\sim0.7$ \citep{Chang-10} and $z\sim0.8$
\citep{Masui-13}.

For redshifts above $z\sim1$, information about atomic gas can  be obtained
through observations of intervening damped $\rm Ly\alpha$ (DLA) or Mg{\,\sc ii}
absorption-line systems in the spectra of background quasars. Surveys of DLAs
and Mg{\,\sc ii} provide an estimate of the total co-moving volume density of
neutral and cool ($\rm 10^4$--$10^5\,K$) gas in the Universe
\citep{Rao-Turnshek-Nestor-06,Rao-17,Prochaska-Wolfe-09,Noterdaeme-12,Crighton-15}.
\cite{Noterdaeme-12} concluded that $\Omega_{\rm HI}$ evolves mildly over the
redshift range $2.3<z<3.5$, while \cite{Zafar-13} showed that no evolution of
$\Omega_{\rm HI}$ is found over the redshift range $1.5<z<5.0$.  A recent study
by \cite{Neeleman-16} at $z\sim$ 0.6 is consistent with  a gradual decline from
$z=2$ to today. \cite{Sanchez-Ramirez-16} found that there is a small but
statistically significant evolution in $\Omega_{\rm HI}$ from $z\sim0$ to
$z\sim5$. More recently \cite{Chowdhury-20}  found that the relation
between H{\,\sc i} mass and absolute blue magnitude does not evolve between
$z=1$ and $z=0$ \citep[see also][]{Walter-20}.

We note that these results are in strong contrast to evolutionary studies
of the CO luminosity function at high redshifts. The cosmic density of
molecular gas in galaxies increases by a factor of 6 from the present day to $z
\sim 1.5$ \citep{Decarli-20}.  This is in qualitative agreement with the
evolution of the cosmic star formation rate density over this redshift
interval, suggesting that the molecular gas depletion time is approximately
constant with redshift. The much weaker trend in $\Omega_{\rm HI}$ with
redshift would argue that the gas traced by H{\,\sc i} is a transient phase,
connected more intrinsically to ongoing gas accetion from the large-scale
environment of the galaxy, rather than to star formation processes occurring in
galactic disks or bulges.

Due to the fact that individual H{\,\sc i} mass measurements are not available
for large galaxy samples at high redshifts, there have been attempts to
calibrate a variety of galaxy properties (usually UV or optical fluxes combined
with stellar mass and other structural parameters) as proxies for the gas to
stellar mass ratio. The H {\,\sc i} gas-to-stellar mass ratio, $M_{\rm
HI}/M_\ast$, has been found to correlate well with, and so can be estimated
from, the optical-optical (e.g. $u-r$) and optical-NIR (e.g. $u-K$) colours
with a typical scatter of $\sim 0.4$ dex \citep[][hereafter K04]{Kannappan-04}.
(We note that, based on higher quality data, a scatter of $\sim$ 0.3 dex has
been found for an estimator of $M_{\rm HI}/M_\ast$ based only on colour by
\citet{Eckert-15}.) After that, some studies have focused on
improving photometric estimators of $M_{\rm HI}/M_\ast$ by defining a
gas-fraction `plane' linking $M_{\rm HI}/M_\ast$, stellar surface mass density,
and optical \citep{Zhang-09} or near-ultraviolet $(NUV)$-optical
\citep{Catinella-10} colour. The scatter in the H{\,\sc i} mass fractions
predicted by these estimators is typically $\sim0.3$ dex in $\log M_{\rm
HI}/M_\ast$. In a more recent study, \citet{Li-12a} further included
$\Delta_{g-i}$, the difference in $g-i$ colour between the outer and inner
region of the galaxy, in order to generate accurate predictions for a
population of massive galaxies with higher-than-average H{\,\sc i} fractions,
which were discovered by \citet{Wang-11} to have bluer, more actively
star-forming outer discs. This estimator is demonstrated to provide unbiased
$M_{\rm HI}/M_\ast$ estimates even for the most H{\,\sc i}-rich galaxies in the
ALFALFA survey. Such photometric estimators have been applied in a number
of recent studies to determine  whether the offset of a galaxy from the mean
mass-metallicity relation depends on its gas content \citep{Zhang-09}, to study
the dependence of galaxy clustering on H{\,\sc i} mass fraction \citep{Li-12a},
\citep{Kauffmann-13}, and to analyze the gas depletion of galaxies in clusters
\citep{Zhang-13}. All these studies are limited to low-redshift galaxies in the
SDSS with $z\la0.3$.

In this paper we extend this previous work by applying a photometric estimator
of $M_{\rm HI}/M_\ast$ calibrated using a local Universe sample of galaxies to a
sample of $\sim7,000$ galaxies with $0.75<z<1.4$ in the DEEP2 survey
\citep{Davis-03}. The estimator is based on the correlation between
$M_{\rm HI}/M_\ast$ and rest-frame $NUV-r$ colour and galaxy size, and is
calibrated using 660 local galaxies that have real H{\,\sc i} emission line
observations.  This provides us with an H{\,\sc i} mass estimate for each galaxy in
our sample, thus  allowing the H{\,\sc i} mass function, the cosmic H{\,\sc i}
mass density and the correlation between atomic gas mass and galaxy properties
such as stellar mass and star formation rate
to be studied for the first time for a large set of galaxies at
$z\sim1$. 
We caution that the conclusions reached in our work rely on the
assumption that the relation between H{\,\sc i} mass fraction and $NUV-r$ colour
and galaxy size in the local Universe applies  at high redshifts. 

The structure of our paper is as follows. In the following section, we describe
the local galaxy sample used for calibrating our H{\,\sc i} mass fraction
estimator, as well as the DEEP2 galaxy sample. Our results are presented in
\S3, where we first present the H{\,\sc i}-to-stellar mass estimator. We then
apply the estimator to the DEEP2 galaxies to study the H{\,\sc i} mass function
(HIMF) (\S3.2) and H{\,\sc i} mass density $\rm\Omega_{\rm HI}$ (\S3.3).  Finally
we investigate the correlations between gas fraction, stellar mass, specific star
formation rate and H{\,\sc i} depletion time in \S3.4.  We summarize our work
in the last section.

Throughout this paper we assume a spatially flat concordance cosmology
with $\Omega_{\rm m}=0.3$, $\Omega_\Lambda=0.7$ and
$H_0=70~\rm km~s^{-1}~Mpc^{-1}$, unless otherwise specified.

\section{Data}
\label{sec:data}

\subsection{The low-z calibrating sample}

We have calibrated the correlation between H{\,\sc i}-to-stellar mass ratio
($M_{\rm HI}/M_\ast$) and $NUV-r$ colour using a sample of 660  local galaxies.
This is a subset of the low-z calibrating galaxies used in \cite{Zhang-09} that
are required to have not only optical data from SDSS and H{\,\sc i} emission
observations from HyperLeda \citep{Paturel-00}, but also imaging in the $NUV$
band from the GALEX surveys \citep{Martin-05}. The derived galaxy parameters
needed for the calibration include H{\,\sc i} gas mass ($M_{\rm HI}$), stellar
mass ($M_\ast$) and rest-frame magnitudes in the $NUV$ and $gri$ bands. The
H{\,\sc i} gas masses are taken from the HyperLeda catalogue following
\cite{Zhang-09}. Following the work of \citet{Wang-10}, we have
reprocessed both the $NUV$-band images from GALEX and the $ugriz$-band images
from SDSS for each of the galaxies to obtain better photometry than available
from the GALEX and SDSS databases. In brief, we first register all the images
to the frame geometry of the NUV image, then convolve the images to a common
PSF of the NUV image, and finally, we perform the photometry measurements
using the same position and aperture as in the SDSS image. We have
corrected all the magnitudes for Galactic extinction, which is determined from
the dust maps of \cite{Schlegel-98} with the extinction curves of
\cite{Cardelli-89} with $R_V=A_V/E(B-V)=3.1$ for the SDSS bands,  implying that
$A_{\lambda}/E(B-V)=8.2$ for the $NUV$ band \citep{Wyder-07}.

\begin{figure}                                                                  
\centering
\resizebox{\hsize}{!}{\includegraphics{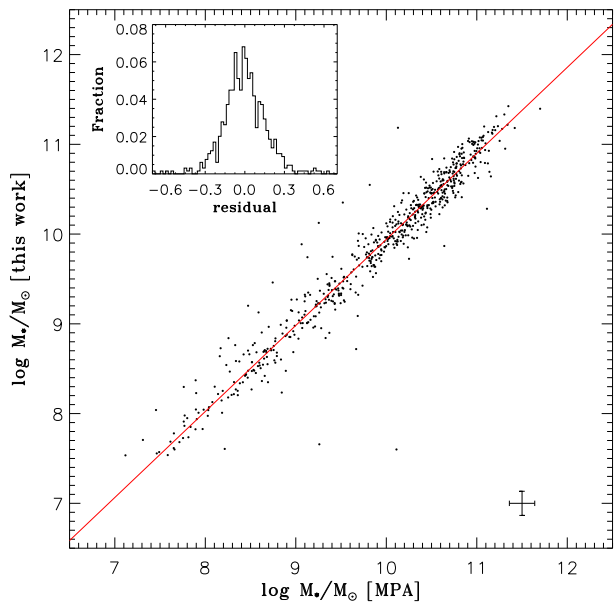}}
\caption{Stellar masses estimated in this work for the galaxies in
the low-z calibrating sample are compared to those from the
MPA/JHU catalogues. The red line represents a linear fit to the data
and the histogram of the residual of galaxies around the best-fit
line is plotted in the inset.}
\label{Fig:compare_local}
\end{figure}

We use a Bayesian approach to estimate a stellar mass from the rest-frame
colours for each galaxy in our sample, following \citet{Kauffmann-03}
and \citet{Salim-05}. A library of 25,000 Monte Carlo realizations of
model star formation histories are generated between $0<z<0.5$ in
regular  bins of $\Delta z=0.001$, using the population synthesis code
of \citet[][BC03]{Bruzual-Charlot-03}. Each star formation history is
characterized with two components: an underlying continuous model with
an exponentially declining star formation law plus random bursts
superimposed on the continuous model.  The models also have
metallicities and dust attenuation uniformly  distributed over wide
ranges. The initial mass function (IMF) of \citet{Kroupa-01}
is adopted. For each galaxy we derive the probability distribution
functions (PDFs) of the stellar mass and rest-frame magnitudes in
different bands by comparing the observed  spectral energy
distribution (SED) to all the model SEDs that are in the closest
redshift interval. In this procedure each model is weighted  by
$\exp(-\chi^2/2)$, where $\chi^2$ is the goodness of fit of the model.
We adopt the mean values of the PDFs as our estimates of these
quantities. The SDSS $ugriz$ SEDs are used when estimating the stellar
masses, while the $NUV$ magnitude is included additionally  when
estimating the rest-frame $g-i$ and $NUV-r$ colours. Our stellar mass
estimates are in good agreement with those from the MPA/JHU SDSS DR4
database\footnote{http://www.mpa-garching.mpg.de/SDSS/DR4/}, as shown
in Fig.~\ref{Fig:compare_local}.

\begin{figure}
\centering
\resizebox{\hsize}{!}{\includegraphics{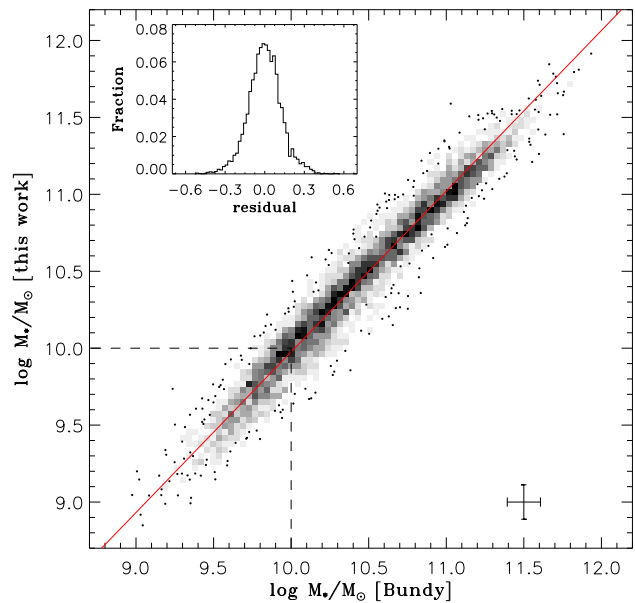}}
\caption{Stellar masses estimated in this work for the DEEP2 galaxies are
compared to those from \citet{Bundy-06}.  The red line represents a linear fit
to the data and the histogram of the residual of galaxies around the best-fit
line is plotted in the inset. \citet{Bundy-06} found that the
spectroscopic sample in DEE2 regions is incomplete below $M_*\sim10^{10}\,\rm
M_\odot$.}
\label{Fig:compare_deep2}
\end{figure}

\subsection{DEEP2 galaxy sample}

The high-z galaxy sample used in this work is based on the third data release
\citep[DR3;][]{Newman-13} of the DEEP2 survey
\footnote{http://deep.berkeley.edu/} and the  $K_{\rm s}$-band selected
catalogue of \citet{Bundy-06} \footnote{http://deep.berkeley.edu/\~{}kbundy/}.
The DEEP2 Galaxy Redshift Survey \citep{Davis-03} utilizes the  DEIMOS
spectrograph \citep{Faber-03} on the Keck II telescope. Targets for the
spectroscopic sample were selected from $BRI$ photometry \citep{Coil-04} taken
with the 12k$\times$8k mosaic camera on the Canada-France-Hawaii Telescope
(CFHT). The images have a limiting magnitude of $R_{\rm AB}\sim25.5$. Since the
$R$-band provides the highest signal-to-noise ratio (S/N) among all the CFHT
bands, the photometry in this band was used to select targets for spectroscopic
observations with the DEEP2 spectrograph. The CFHT imaging covers four widely
separated regions, with a total area of 3.5 deg$^2$. In fields $2-4$, the
spectroscopic sample is pre-selected using $(B-R)$ and $(R-I)$ colours to
eliminate objects with $z<0.7$ \citep{Davis-03}. Colour and apparent magnitude
cuts were also applied to objects in the first field, the Extended Groth Strip
(EGS), but these were designed to downweight low-redshift
galaxies to select roughly equal number of galaxies below and above $z=0.7$
\citep{Willmer-06}.

Based on the DEEP2 sample, \citet{Bundy-06} conducted an extensive
imaging survey of all the DEEP2 fields with the Wide Field Infrared
Camera \citep[WIRC;][]{Wilson-03} on the 5-metre Hale Telescope at
Palomar Observatory. Using contiguously spaced pointings, the central
third of fields $2-4$ was mapped to a median $80\%$  completeness depth
greater than $K_{\rm AB}=21.5$, accounting for $0.9$ deg$^2$ on the sky. 
The EGS field covers $0.7$ deg$^2$ with varying but deeper
depths. A stellar mass was estimated by \citet{Bundy-06} for each of
their galaxies. First, a $K_{\rm s}$-band mass-to-light ratio
$M_*/L_{K_{\rm s}}$ was obtained by comparing the observed $BRIK_{\rm s}$ SED  to
a grid of 13,440 synthetic SEDs constructed from BC03 spanning a
range of star formation histories, ages, metallicities and dust
content. The stellar mass of the galaxy is then given by scaling 
$M_*/L_{K_{\rm s}}$ to the $K_{\rm s}$-band luminosity measured from the total
$K_{\rm s}$-band magnitude and the DEEP2 spectroscopically-measured
redshift. 

In this work, we make use of the galaxy sample from \citet{Bundy-06},
which consists of 7,222 galaxies with redshifts in the range  $0.75\le
z \le 1.40$, redshift quality parameters $z_{\rm quality} \ge 3$ and
$K_{\rm s}$-band magnitudes $K_{\rm s} \le 20$. For consistency, we estimate  a
stellar mass for each of the galaxies using the same method we  used
above for the low-z calibrating sample. The observed SED used  for the
estimation involves photometry in $B$, $R$, $I$ and $K_{\rm s}$ bands.  Our
stellar mass estimates are compared to those obtained by
\citet{Bundy-06} in Fig.~\ref{Fig:compare_deep2}, which are in good
agreement with no obvious systematic difference.  We have also
estimated the rest-frame $g-i$ and $NUV-r$ colours for the galaxies
using the same method.

\begin{figure}
\centering
\resizebox{\hsize}{!}{\includegraphics{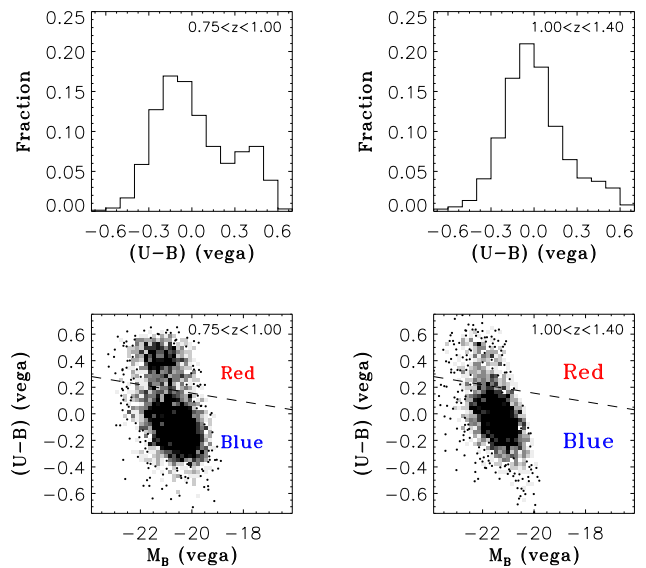}}
\caption{Histograms of the Rest-frame $U-B$ colour (top panels)
  and diagrams of $U-B$ vs. $B$-band absolute magnitude (lower panels)
  are shown for DEEP2 galaxies in two successive redshift intervals:
  $0.75<z<1$ (left panels) and $1<z<1.4$ (right panels).
  The dashed line in the lower panels is the luminosity-dependent
  colour divider employed by \citet{vanDokkum-08}, which is used in
  this work to classify our galaxies into red and blue populations.
  Magnitudes plotted in this figure are in the Vega system.}
\label{Fig:redblue}
\end{figure}

Following previous studies \citep[e.g.][]{Bundy-06, Chen-09} we weight
each galaxy in our DEEP2 sample to correct for incompleteness when
performing statistical analyses:
\begin{equation}
W=\frac{\kappa}{V_{\rm max}}, 
\end{equation}
where $\kappa$ accounts for incompleteness resulting from the  DEEP2
colour selection and redshift success rate. The ``optimal" weighting
model of \citet{Willmer-06} is used to estimate $\kappa$, which
accounts for the redshift success rate for red and blue galaxies  in
different ways. We adopt the luminosity-dependent colour divider
employed by \cite{vanDokkum-08} for classifying the galaxies into red
and  blue:
\begin{equation}\label{eqn:colour_divider}
U-B = -0.032(M_{\rm B} + 21.52) + 0.454 - 0.25,
\end{equation}
where the rest-frame $U-B$ colour and the $B$-band absolute magnitude
$M_{\rm B}$ are obtained above using our Bayesian approach. The $U-B$
histogram and the $M_{\rm B}$ versus $U-B$ diagram are shown in
Fig.~\ref{Fig:redblue}, for two successive redshift intervals:
$0.75<z<1$ and $1<z<1.4$. 

The second factor $V_{\rm max}$ is defined as the maximum volume over which the
galaxy would be included in the sample, accounting for the fact that faint
galaxies are not detected throughout the entire survey volume in a flux-limited
survey. Following \citet{Willmer-06}, the $V_{\rm max}^i$ for a
galaxy $i$ can be calculated as:
\begin{equation}
V_{max}^i = \int_{\Omega}\int_{z_{min,i}}^{z_{max,i}}\frac{d^2V}{d\Omega dz}dzd\Omega
\end{equation}
where $z$ and $\Omega$ are the redshift and the solid angle, respectively. The
redshift limits $z_{min,i}$ and $z_{max,i}$ are imposed either by the limits of
the redshift bin being dealt with or by the apparent magnitude limits of the
DEEP2 sample. More details can be found in \citet{Willmer-06}.

\begin{figure}
\centering
\resizebox{\hsize}{!}{\includegraphics{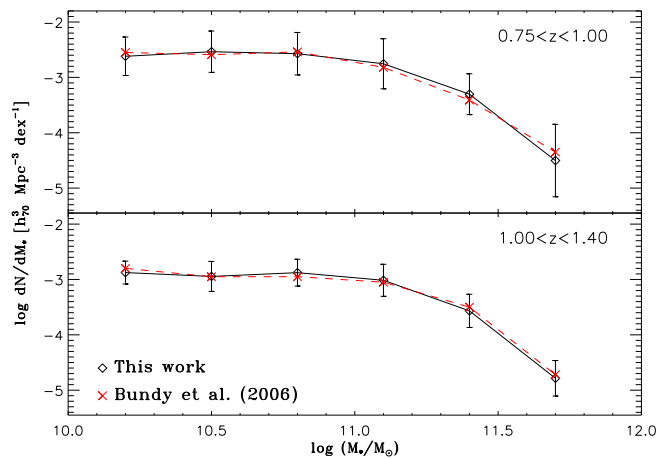}}
\caption{Stellar mass functions at the two redshift intervals as 
  indicated are plotted in black diamonds, and are compared to 
  the measurements from \citet{Bundy-06} plotted in red crosses.}
\label{Fig:MF}
\end{figure}

We have estimated stellar mass functions for the two redshift bins, which are
plotted in Fig.~\ref{Fig:MF} as diamonds connected by solid lines, and these are
compared to the estimates obtained by \cite{Bundy-06}, shown as crosses
connected by dashed lines. The two estimates are almost identical, indicating
that we have correctly calculated the weights for incompleteness correction.
For the mass function estimates (and the H{\,\sc i} mass functions in the next
section) we have used 5,372 out of the 7,222 galaxies with
$M_\ast>10^{10}$\,$\rm M_\odot$. Below this mass, the DEEP2 sample
becomes significantly incomplete \citep{Bundy-06}. The errors on the stellar
mass functions are estimated using the $1\sigma$ scatter in the
measurements of the four DEEP2 fields. 

For one of the four DEEP2 fields, the Extended Groth Strip
(EGS),  SFR measurements of galaxies have been carried out by
\cite{Barro-11} using SEDs covering the wavebands from UV to FIR, including the
IRAC 3.5 and 4.6 $\mu m$ bands. About 2000 blue galaxies in the EGS field in
our sample have SFR measured in this way.

\begin{figure*}
\centering
\includegraphics{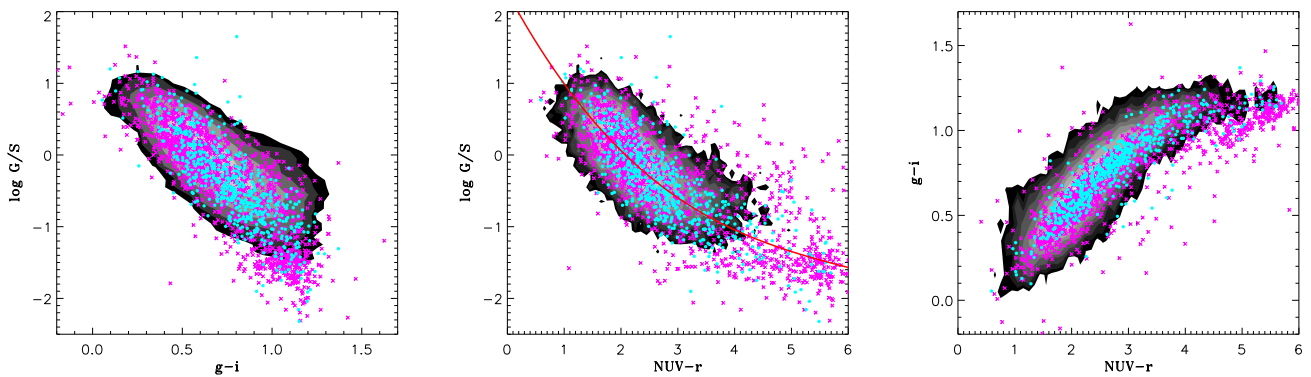}
\caption{Correlation of H{\,\sc i}-to-stellar mass ration ($M_{\rm HI}/M_\ast$)
with $g-i$ (left) and $NUV-r$ (middle) for the low-z calibrating sample. The
best-fit models are plotted as a solid red line (see the text for detailed
description). The right-hand panel shows the relation between $g-i$ and $NUV-r$
for the same set of galaxies. The low-z calibrating sample are
plotted as cyan filled dots, the RESOLVE sample are plotted as magenta crosses,
while the the final ALFALFA 100 per cent catalogue ($\alpha$.100) are shown as
black-and-white contours in each panel.}
\label{Fig:estimator}
\end{figure*}

\section{Results}

\subsection{The H{\,\sc i} mass fraction estimator}
\label{sec:estimator}

The H{\,\sc i}-to-stellar mass ratio is strongly correlated with galaxy
parameters such as optical colours, $NUV-r$ colour, $\mu_*$, $R_{90}/R_{50}$
and $M_\ast$, with a typical scatter of $\sim 0.35$ dex. The relation between
$M_{\rm HI}/M_\ast$ with the combination of two or more parameters is tighter
than the relation between $M_{\rm HI}/M_\ast$ with a single parameter, with a
typical scatter of $\sim 0.30$ dex \citep{Zhang-09,Li-12a,Catinella-10}. Unlike
\cite{Zhang-09}, we take the combination of $NUV-r$ and $\mu_*$, instead of
optical colour and $\mu_*$ in this work, because of two considerations: 1) As
we see from Fig.~\ref{Fig:estimator}, low redshift galaxies show very narrow
dynamic range in $g-i$ with $1\la g-i\la 1.2$, but $NUV-r$ spans a broader
range of values ($3.5\la (NUV-r)\la 6$). This is more clearly seen in the
right-hand panel of the same figure, where $g-i$ is plotted against $NUV-r$.
When $NUV-r$ increases above $\sim3.5$, $g-i$ doesn't increase any more and is
fairly constant at $g-i\approx1.1$, suggesting that $NUV-r$ is  more sensitive
than $g-i$ to the atomic gas content for the most H{\,\sc i}-poor galaxies.
This is consistent with the finding of \citet{Catinella-10} that, among the
four parameters considered ($M_\ast$, $\mu_\ast$, $R_{90}/R_{50}$, $NUV-r$),
the $NUV-r$ colour was the one most tightly correlated with the H{\,\sc
i}-to-stellar mass ratio.  2)for the DEEP2 galaxies at $z\sim1$, their
rest-frame $NUV$ magnitude is redshifted to the optical band, and in this case
the $k$-correction is relatively small and can be estimated more reliably than
in the case where the $g$ magnitude is used. 

The physical reason why H{\,\sc i} is better predicted by NUV-r colour likely
arises from the fact that the ultra-violet emission from a galaxy traces the
light from young stars in low density, extended  gas in the outer disks of
galaxies that has not been absorbed by dust. The molecular gas traced by CO
emission is, on the other hand, known to be tightly correlated with the
infrared luminosity of the galaxy \citep{Young-84,Sanders-85}. This analysis
will not address the molecular content of the galaxies in our sample.

We have checked the results of the low-z calibrating sample using more recent
data from two larger H{\,\sc i} catalogues.  The first one is the REsolved
Spectroscopy of a Local VolumE (RESOLVE\footnote{http://resolve.astro.unc.edu})
sample, which is completed in baryonic mass (defined as $M_{bary}=M_*+1.4M_{\rm
HI}$) down to dwarfs of~$\sim10^9$\,$\rm M_\odot$ \citep{Eckert-15,Stark-16}.
The second one is the final ALFALFA 100 per cent catalogue
($\alpha.100$\footnote{http://egg.astro.cornell.edu/alfalfa/})\citep{Jones-18,Haynes-18}.
The galaxies with signal-to-noise lower than 5 are excluded from this sample.
The RESOLVE and the $\alpha.100$ samples are shown in Fig.~\ref{Fig:estimator}
as magenta crosses and black-and-white contours, respectively.  Note that for
these two samples, the colours and the stellar masses used here are obtained
from the NASA-Sloan Atlas (NSA) catalogue
(v1\_0\_1\footnote{https://data.sdss.org/sas/dr16/sdss/atlas/v1/}), and the
colours are corrected for Galactic extinction. As a result, about 1,500 RESOLVE
galaxies and about 13,000 $\alpha.100$ galaxies are used to compare with the
low-z calibrating sample. We can see that although the low-z calibrating sample
is small, it follows the same trend with the RESOLVE and the $\alpha.100$
samples.  This confirms that the estimator derived using low-z calibrating
sample is robust.

We use a polynomial function to  represent the observed nonlinear
relation of $M_{\rm HI}/M_\ast$ versus $NUV-r$, 
\begin{eqnarray}
\rm \log G/S & = & a_0 + a_1 (NUV-r)+ \nonumber \\ &   & a_2
(NUV-r)^2 + a_3 (NUV-r)^3,
\label{Eq:estimator}
\end{eqnarray}
plotted as a red line in the middle panel of
Fig.~\ref{Fig:estimator}. Here the best-fit parameters are
$a_0=2.26432$, $a_1=-1.46308$, $a_2=0.202311$ and $a_3=-0.0108242$.
The $1\sigma$ scatter of the galaxies around this relation is $0.36$
dex.

\begin{figure}
\centering
\resizebox{\hsize}{!}{\includegraphics{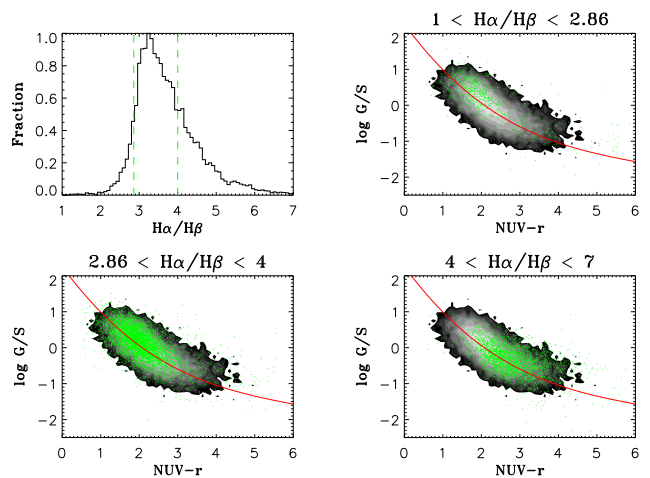}}
\caption{The effect of dust attenuation to the relation between $M_{HI}/M_*$
and $NUV-r$ colour. The distribution of H$\alpha$-to-H$\beta$ ratio of the
$\alpha.100$\_spec sample is shown in the top-left panel. The green dashed
lines indicate the division of the sample into three subsamples. In the other
panels, the three subsamples are shown as green dots in $M_{HI}/M_*$ vs.
$NUV-r$ diagram.  Similar to the middel panel in Fig.~\ref{Fig:estimator}, the
black-and-white contours are the final ALFAFA 100 per cent catalogue
($\alpha.100$), while solid red line is the relation described in Eq.
\ref{Eq:estimator}. As can be seen, galaxies of different dust content are
shifted along the derived relation, not displaces away from it.}
\label{Fig:HaHb_ratio}
\end{figure}

We have explored possible systematic effects due to dust
attenuation on the correlation between H{\,\sc i} gas fraction and $NUV-r$
colour. H$\alpha$-to-H$\beta$ decrement can be used as an indicator of dust
attenuation, for which higher values of H$\alpha$/H$\beta$ indicate higher dust
attentuation values. We obtain a sample of $\sim$ 10,000 galaxies (hereafter
$\alpha.100$\_spec sample) which have high signal-to-ratio ($S/N>3$) detection
of H$\alpha$ and H$\beta$ emission lines, by cross-matching the $\alpha.100$
sample with SDSS DR8 spectroscopic sample. We note that all galaxies
on the star-forming main sequence in SDSS have H$\alpha$ and H$\beta$ 
with this $S/N$ or higher. We divide the sample into three
subsamples according to the H$\alpha$-to-H$\beta$ ratio as shown in the
top-left panel in Fig.~\ref{Fig:HaHb_ratio}, and we plot these subsamples in the
H{\,\sc i} gas fraction and $NUV-r$ colour diagram in the other three panels.
As in the middle panel of Fig.~\ref{Fig:estimator}, the final ALFAFA 100 per cent
catalogue ($\alpha.100$) is shown as black-and-white contours for comparison.
The galaxies having H$\alpha$ and H$\beta$ detections are over-plotted as green
dots, with the cuts noted on the top of each panel. The solid red line is the
relation described in Eq. \ref{Eq:estimator}. It is found that the galaxies
with higher dust attenuation have redder $NUV-r$ colours and lower H{\,\sc i}
gas fractions. However, they still follow the correlation between H{\,\sc i}
gas fraction and $NUV-r$ colour. We conclude that it is safe to apply the
estimator to galaxies with various dust attenuation values.

\begin{figure}
\centering
\resizebox{\hsize}{!}{\includegraphics{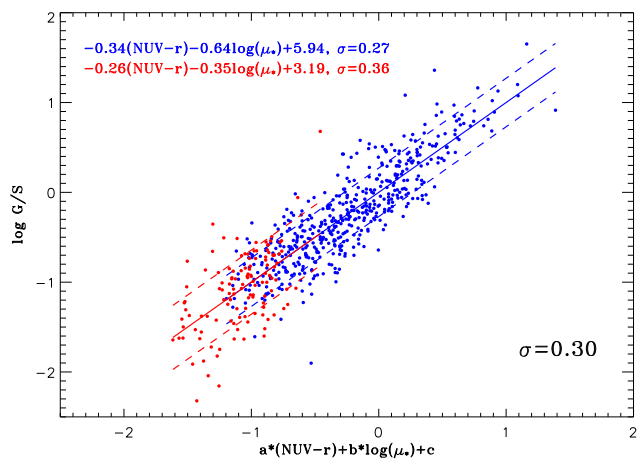}}
\caption{The relation between H{\,\sc i} gas fraction and the linear
combination of $NUV-r$ colour and stellar surface mass density ($\mu_*$) in the
SDSS-z band. The low-z calibrating sample has been divided into two subsamples
according to $NUV-r$ colour: the red points are galaxies with $(NUV-r)>3.5$,
while the blue points are galaxies with $(NUV-r)<3.5$. The fitting parameters
are labeled at top-left of the figure.
}
\label{Fig:estimator_nuvr_mustar}
\end{figure}

Fig.~\ref{Fig:estimator_nuvr_mustar} shows the relationship between H{\,\sc
i}-to-stellar mass ratio, $G/S$, and a linear combination of $NUV-r$ and
stellar surface mass density $\mu_*$, defined as 0.5$M_*/R_{50}^2$, where $M_*$
is the stellar mass of the galaxy and $R_{50}$ is its z-band half light radius
in units of kpc, for the galaxies in the low-z calibrating sample. This
estimator could be expected to exhibit less scatter than one based on $NUV-r$
colour alone if the surface density of the atomic gas is well correlated with
the surface density of the stars in galaxies.  In practice, the H{\,\sc i}
often extends far beyond the disk traced in the optical light. \cite{Wang-13}
show that there is a factor of $\sim 1.5$ scatter in ratio between H{\,\sc i}
size and optical size at fixed stellar mass, inplying a factor of 2 or more
scatter in surface density. We find that the estimator that includes $\mu_*$
has a scatter around the one-to-one line of 0.30 dex.  We can obtain somewhat
tighter results if we fit separate relations for galaxies with $(NUV-r)>3.5$
and $(NUV-r)<3.5$.  The scatter of blue and red galaxies around their separate
relation  is 0.27 and 0.36 dex, respectively. The higher scatter for redder
galaxies likely occurs because these galaxies contain more dust and a higher
fraction of their total gas content is in molecular rather than atomic form.

\begin{figure}
\centering
\resizebox{\hsize}{!}{\includegraphics{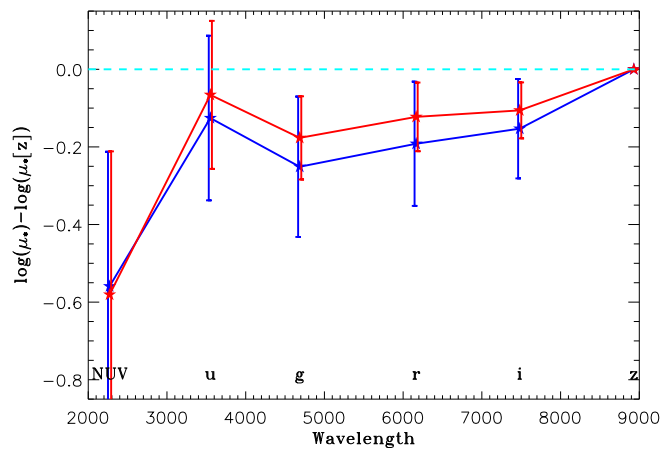}}
\caption{The difference of $\mu_*$ in GALEX-$NUV$, sdss-$u,g,r,i$ bands with
that in sdss-z band. As in Fig.~\ref{Fig:estimator_nuvr_mustar}, the red and
blue colours represent galaxies with $(NUV-r)>3.5$ and $(NUV-r)<3.5$,
respectively.}
\label{Fig:diff_mustar}
\end{figure}

\begin{figure}
\centering
\resizebox{\hsize}{!}{\includegraphics{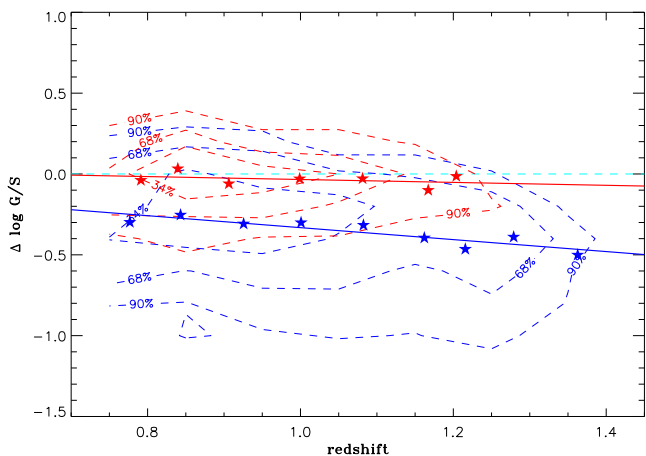}}
\caption{The difference between G/S estimated from $NUV-r$ and $\mu_*$ and G/S
estimated from $NUV-r$ only as function of redshift. The median value in
each redshift bin and the linear fitting results of these median values are
shown as star and solid line, respectively. The contours enclose 34, 68, and
90\% of the sample. Similar to Fig.~\ref{Fig:estimator_nuvr_mustar}, the red
and blue colour represent galaxies with $(NUV-r)>3.5$ and $(NUV-r)<3.5$,
respectively.}
\label{Fig:compare_new_old_GS_z}
\end{figure}

This estimator cannot be applied into the whole high-z samples directly, because
only one of the four DEEP2 fields, the EGS region, has half-light radius
determined for each galaxy from HST observations.  Before we apply the
estimator to the EGS galaxies, we notice the fact that the HST half-light
radius is determined in the $I$-band with effective wavelength of 8140\,\AA,
corresponding to a rest-frame wavelength ranging from 3391\,\AA~to 4651\,\AA~for our
galaxies. However, the $\mu_*$ for the low-z calibrating sample is obtained from the
SDSS $z$-band at 8931\,\AA. As shown in Fig.~\ref{Fig:diff_mustar}, the surface
mass density $\mu_\ast$ of galaxies changes from band to band due to the
wavelength dependence of their half-light radii. In order to take into account
this effect, we use the median difference in $\log\mu_\ast$ shown in the figure
to make a correction to $\mu_\ast$ for galaxies in the EGS.  Then for galaxies
in the EGS region, we compare the gas fractions estimated from the linear
combination of $NUV-r$ and $\mu_*$ with the gas fractions given by $NUV-r$
only, and find systematic differences that depend on redshift as shown in
Fig.~\ref{Fig:compare_new_old_GS_z}.  For red and blue galaxies, the gas
fraction taking into account $\mu_*$ is smaller by 0$-$0.1 and 0.2$-$0.5 dex,
respectively. 

In summary,in this section we have demonstrated that among the quantities
we have access to, the combination of $NUV-r$  colour stellar surface density
yields the best  predictor for H{\,\sc i} mass fraction for blue galaxies in
particular.  Use of a single colour such as $NUV-r$ leads to inaccurate
predictions, depending on the weighting between red and blue galaxies in the
sample.  We show that a separate calibration for the two populations is
necessary to obtain accurate estimates of the high-mass end of the HIMF and the
cosmic H{\,\sc i} mass density. In this analysis, we apply the median
difference, shown in the red/blue stars and lines in the Fig.~
\ref{Fig:compare_new_old_GS_z}, to correct the gas fraction for all the
galaxies in our DEEP2 sample. We note that our analysis in the following
sections does not include any estimate of the molecular gas content of the
galaxies and this component may have very different scaling relations than the
atomic gas.

\subsection{H{\,\sc i} mass function at $z\sim1$}

\begin{figure}
\centering
\resizebox{\hsize}{!}{\includegraphics{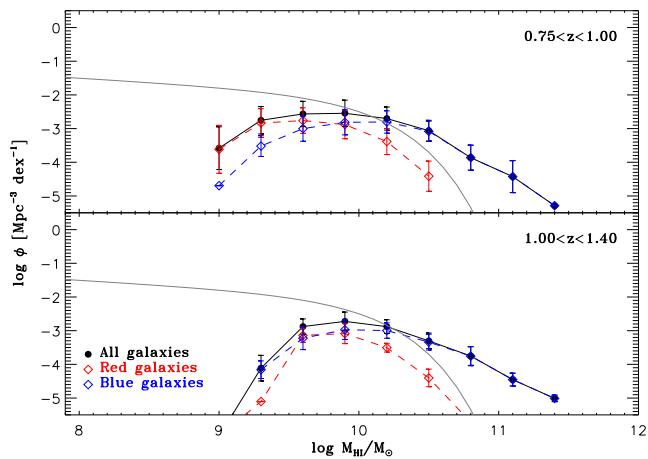}}
\caption{ H{\,\sc i} mass functions estimated from the DEEP2 for
  the two redshift intervals as indicated are plotted in black solid
  circles connected by a solid line for the whole sample,
  and in red (blue) diamonds connected by a dashed line for
  the red (blue) galaxy subsamples. For comparison, the
  HIMF at $z=0$ from $\alpha.100$ is shown as grey solid
  lines.
}
\label{Fig:himf_new}
\end{figure}

Applying the H{\,\sc i} mass fraction estimator obtained above, we have
estimated  an H{\,\sc i} gas mass for each of the galaxies in the DEEP2 sample.
The first  statistic that we have obtained from this data is the H{\,\sc i}
mass function, which quantifies the co-moving number density of galaxies as a
function of their H{\,\sc i} mass. This statistic has not been estimated
previously at these high redshifts. Specifically, we have estimated the HIMF
for two successive redshift intervals, one below and one above $z=1$. These are
shown in Fig.~\ref{Fig:himf_new} as black solid circles connected by a solid
line. When estimating the HIMFs, we have corrected for sample incompleteness in
the same way as described above for estimating the stellar mass functions. We
have restricted ourselves to galaxies with log($M_*$/$\rm M_{\odot})>10$ for the
same reason as explained above. The errors on the HIMFs are estimated from the
$1\sigma$ scatter of the HIMFs of the four separate fields of the DEEP2. 

For each redshift interval, we have also estimated an HIMF for the subset of
galaxies with either red or blue $U-B$ colours, classified using the colour
divider in equation~(\ref{eqn:colour_divider}). The estimates are also shown in
Fig.~\ref{Fig:himf_new}.  For comparison, the HIMF of the local Universe
obtained by \citet{Jones-18} from the final catalogue of the ALFALFA
($\alpha$.100)  is plotted as a grey solid line in the same figure. 

We see that at H{\,\sc i} masses above $\sim10^{10}$\,$\rm M_\odot$, the abundance of
galaxies decreases slowly with increasing H{\,\sc i} mass before it drops
rapidly at masses above $\sim10^{11}$\,$\rm M_\odot$. This behaviour is similar to the
luminosity functions and stellar mass functions found in the local Universe,
which can usually be described by a Schechter function. At masses below
$\sim10^{10}$\,$\rm M_\odot$, the HIMFs also decreases, and this should be attributed
to the fact that the galaxies with stellar mass below $10^{10}$\,$\rm M_\odot$ are not
included in the measurement. For this reason, the HIMFs presented here can only
be regarded as lower limits of the number density of galaxies with 
H{\,\sc i} masses lower than $\sim10^{10}$\,$\rm M_\odot$. Hence in the following
analysis, we only fit a Schechter function to the HIMFs at  H{\,\sc i} masses
above $10^{10}$\,$\rm M_\odot$.

\begin{figure}
\centering
\resizebox{\hsize}{!}{\includegraphics{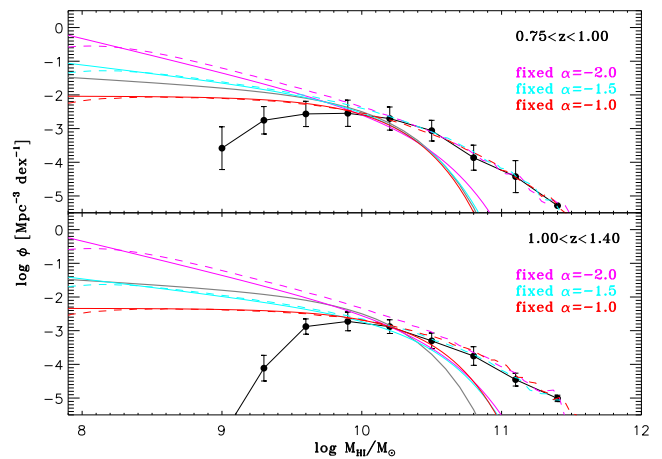}}
\caption{Schechter function fits with the slope fixed to $\alpha=-2,-1.5,-1.0$
are shown as magenta, cyan and red dashed lines. During fitting, a 0.35 dex
scatter of $\rm log(M_{\rm HI})$ is adopted. The corresponding Schechter
functions without scatter in $\rm log(M_{\rm HI})$ are shown as solid lines.  }
\label{Fig:himf_scatter}
\end{figure}

We note that the errors on the H{\,\sc i} masses can affect the shape of the
HIMFs, in particular at the high mass end. Given that the scatter in our
estimates of the gas fraction log($M_{\rm HI}/M_*$) and in the stellar mass
log(M$_*$) is 0.30 dex (see Fig.~\ref{Fig:compare_deep2}) and 0.11 dex (see
Fig.~\ref{Fig:estimator_nuvr_mustar}), the scatter of H{\,\sc i} gas mass
log($M_{\rm HI}$) is estimated as 0.32 dex.  Considering the uncertainties  of the
corrections described in \S\ref{sec:estimator}, a total error in log($M_{\rm HI}$)
of 0.35 dex is assumed. We then fit a Schechter function to the HIMFs at
H{\,\sc i} masses above $10^{10}$\,$\rm M_\odot$, by considering the scatter of $\rm
log (M_{\rm HI})$ with value of 0.35 dex. We note that the flat slope of the HIMF
at low masses in highly uncertain because of the extrapolation to low masses.
We have tried to fit to the HIMF measurements with the slope parameter fixed to
three different values: $\alpha=-1$,  $-1.5$ and $-2$, and plot the best-fit
functions as dashed red, cyan and magenta lines in Fig.~
\ref{Fig:himf_scatter}. As can be seen from the figure, fitting to the HIMFs
with different Schechter function slopes cannot effectively constrain the slope
parameter at the low-mass end. At the high-mass end, the shape is well
constrained, and consistent with no evolution at the bright end.

\subsection{H{\,\sc i} mass density at $z\sim1$}

The second quantity that we have obtained from our data is the cosmic H{\,\sc
i} mass density, $\Omega_{\rm HI}$, which is the volume density of H{\,\sc i}
mass contained in galaxies at given redshift normalized by the critical density
of matter in the universe.  Because of large uncertainties for the slope at the
low-mass end, we did not try to estimate the total $\Omega_{\rm HI}$ from the
best-fit Schechter functions shown in Fig.~\ref{Fig:himf_scatter}. Instead, we
measure the lower limits of $\Omega_{\rm HI}$ by directly integrating the
HIMFs, using galaxies with $M_*>10^{10}$\,$\rm M_\odot$. The
scatter among the four DEEP2 fields is adopted as the
1$\sigma$ uncertainty on the measurement. We derive lower limits $\Omega_{\rm
HI, limit}$  $(2.44 \pm 0.76)\times10^{-4}\, h_{70}^{-1}$ and
$(1.66\pm0.54)\times10^{-4}\, h_{70}^{-1}$ for the two redshift bins.  We note
that the mean measurement of $\Omega_{\rm HI, limit} \sim 2.1 \times10^{-4}\,
h_{70}^{-1}$ at $z\sim1$ is comparable with that of $\Omega_{\rm HI, Bright} =
(2.31 \pm 0.58) \times10^{-4}\, h_{70}^{-1}$ by directly stacking of 21-cm
signatures of blue, star-forming galaxies with $M_{\rm B} \leq -20$ at similar
redshifts in the DEEP2 fields \citep{Chowdhury-20}.

\subsection{Correlations between H{\,\sc i} mass fraction, stellar mass, specific star formation rate, and H{\,\sc i} depletion time}

\begin{figure} \centering \resizebox{\hsize}{!}{\includegraphics{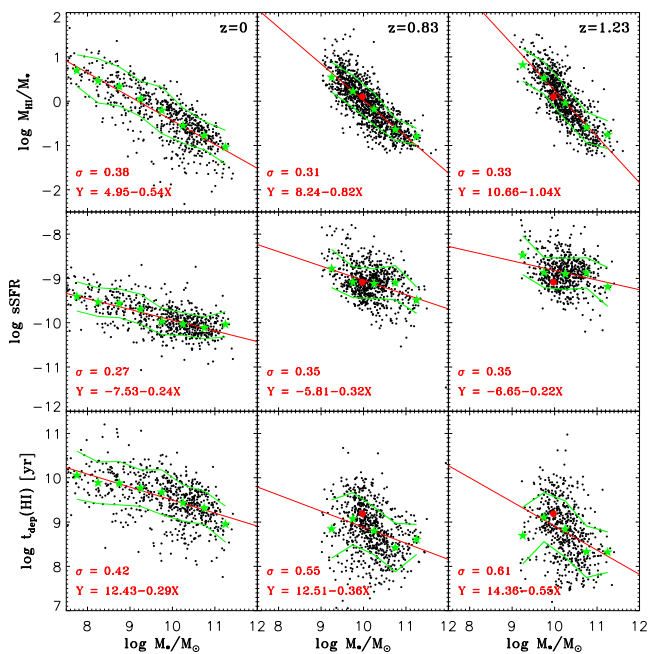}}
\caption{The H{\,\sc i}-to-stellar mass ratio ($M_{\rm HI}/M_\ast$; top panels),
specific star formation rate (SFR/$M_\ast$; second row of panels) and the H{\,\sc
i} gas depletion time ($M_{\rm HI}/$SFR; bottom panels) are plotted as a function
of stellar mass, for the low-z calibrating sample (left) and the DEEP2
galaxies at the two higher redshifts (center and right panels). The small black
dots are for the low-z calibrating sample or the blue galaxies at high redshifts, 
and the red line in each panel shows the best fitting with a linear relation.
The stars and the two solid lines plotted in green show the median and
$1\sigma$ scatter for a number of stellar mass bins. For comparison, the mean
measurements of star-forming galaxies at redshift $z\sim1$ by
\cite{Chowdhury-20} are plotted as red filled circles.}
\label{Fig:Mass_etal}
\end{figure}
 
\begin{figure}
\centering
\resizebox{\hsize}{!}{\includegraphics{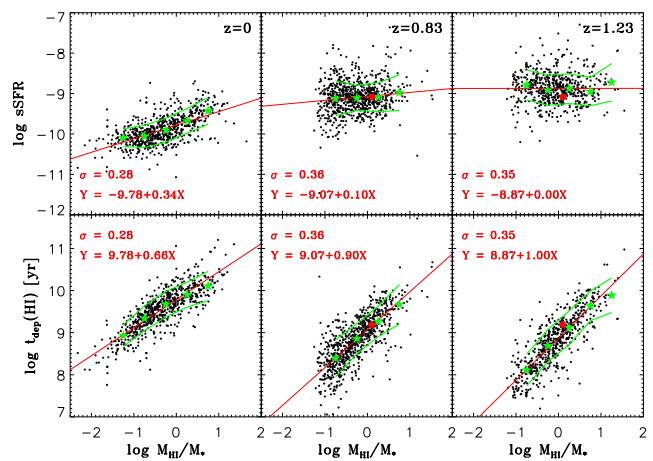}}
\caption{Correlations of specific star formation rate and H{\,\sc i} gas
depletion time with H{\,\sc i}-to-stellar mass ratio are plotted for the
same sets of galaxies with the same symbols/lines as in the
previous figure.
  }
\label{Fig:GS_etal}
\end{figure}

In this subsection we study the correlation of the H{\,\sc i} gas content of
the galaxies at $z\sim1$ with their star formation rate (SFR) using galaxies
from the the Extended Groth Strip (EGS), for which SFR measurements of galaxies
are available. The SFR measurements range from 1 to 3 $\rm M_\odot\,yr^{-1}$,
with  a median value of 1.2  $\rm M_\odot\,yr^{-1}$.  In
Fig.~\ref{Fig:Mass_etal} and~\ref{Fig:GS_etal} we use this sample (again, split
into two subsamples by redshift), as well as the low-z calibrating sample, to
examine the correlations between the specific SFR, the H{\,\sc i} mass
fraction, and the quantity $M_{\rm HI} / {\rm SFR}$. We note that $M_{\rm HI} /
{\rm SFR}$ should be regarded as a lower limit to the actual time taken for the
H{\,\sc i} gas to be depleted. In low redshift galaxies, most of the
H{\,\sc i} gas is often well outside the star-forming disk in a dynamically
stable configuration. Some of this gas may be  a reservoir for future star
formation in the galaxy, but the timescale for consumption of the gas will
depend on uncertain parameters such as gas inflow rates.  At high redshifts,
gas inflow rates are higher, the dynamical timescales of galaxies are shorter
and extended gas reservoirs  may be converted into stars more quickly. Therefor
one might expect to see better correlation between the star formation rate and
the H{\,\sc i} gas content at high redshift.  Our results,however,indicate that
this is not true.

The top three panels in Fig.~\ref{Fig:Mass_etal} show the distribution of
galaxies in the $\log M_{\rm HI}/M_\ast$ versus $\log M_\ast$ plane.  The low-z
calibrating sample shows a tight, nearly-linear relation with a slope of
$-0.54$ and a $1\sigma$ scatter of 0.38 dex. A similarly tight relation is seen
at $z=0.83$ and $z=1.23$, with scatter 0.31 and 0.35 dex.  The correlation
seems to steepen to higher redshifts; fitting to  the data with a linear
relation leads to slope parameters of -0.82 and -1.04 for the two high-z
samples compared to $-0.54$ for the low-z sample. To take into account the
effect of incompleteness, we also fit the data weighted by $1/V_{max}$.  The
resulting slopes are  $-0.72$ and $-0.93$.  The normalization of the
correlation also increases with redshift.  At stellar mass of $10^{11}$\,$\rm
M_\odot$ the average H{\,\sc i}-to-stellar mass ratio  is $0.1$ at $z\sim0$,
and this increases by a factor of 3--4, up to $\sim0.3$ at $z\sim0.8$ and
$\sim0.4$ at $z\sim1.2$. The increase is even more dramatic at smaller stellar
masses, e.g.  by a factor of 4--12 at $M_\ast=10^{10}$\,$\rm M_\odot$. These
results show that distant galaxies are on average much more H{\,\sc i}-rich
than local galaxies, and the slope of the relation between H{\,\sc i} gas mass
fraction and stellar mass also evolves with redshift.  We have checked the
nature of the galaxies with $M_{\rm HI} > 10^{10}$ M$_{\odot}$ in the EGS
field, and have found that their half-light radii in HST I-band vary from 20 to
90 kpc, and the surface mass density of H{\,\sc i} gas using this this size
estimate is about 37 $\rm M_{\odot}\,pc^{-2}$.

In the middle panels we reproduce the star formation sequence by showing the
log(SFR/$M_\ast$) versus $\log M_\ast$ planes. There is a similar increase in
the normalization of the relation. Interestingly, the relation between sSFR and
stellar mass shows larger scatter at $z\sim1$ than the relation between
H{\,\sc i} gas mass fraction and stellar mass. We note that the molecular gas
fraction also increases by a factor of 3$-$4, from $z=0$ to $z\sim0.8$
and $z\sim1.2$, according to the evolution trend of
$f_{\rm mol-gas}\propto(1+z)^2$ \citep{Geach-2011,Carilli-2013}. 

In the bottom panels of Fig.~\ref{Fig:Mass_etal}, we show the correlation
of  $M_{\rm HI} / {\rm SFR}$    with stellar mass for the same sets of
galaxies. This quantity yields rather short  timescales at all redshifts probed
(0.5$-$2 Gyr at $z\sim1$ and a factor only $\sim$2 larger at $z\sim0$), but
aswe have noted, these are likely lower limits to the actual time taken for the
gas to be consumed. More interesting is that this quantity exhibits very weak
redshift evolution, similar to the H{\,\sc i} gas and star formation main
sequences  discussed above. This indicates that the star formation main
sequence may simply be a natural result of the cosmic evolution of the
available gas reservoir in galaxies.  Furthermore, it is also seen that $M_{\rm
HI} / {\rm SFR}$    at fixed stellar mass shows a huge scatter, $\sim0.6$ dex
at $z\sim1$ and $\sim0.4$ dex in the local Universe, implying a weak
correlation between $M_{\rm HI}$ and SFR, as expected for a gas component that
is not directly linked with the star formation in the galaxy population.

In the upper panels of Fig.~\ref{Fig:GS_etal}, 
we plot the galaxies in the plane of log(SFR/$M_\ast$) versus $M_{\rm HI}/M_\ast$.
At $z\sim1$ the sSFR is confined to a narrow range, again reflecting the
tightness of the star formation sequence, while $M_{\rm HI}/M_\ast$ spans a much
wider range. This causes the quantity $M_{\rm HI} / {\rm SFR}$  to be tightly
correlated with $M_{\rm HI}/M_\ast$, as shown in the lower panels of the same
figure. 

The scaling relations shown in Figures 13 and 14 are best interpreted in
conjunction with either semi-analytic models of galaxy formation that model the
spatial distribution of the gas
\citep[e.g.][]{Fu-10,Lagos-11,Popping-14,Xie-17} or cosmological hydrodynamical
simulations with sufficient (1 kpc or better) resolution to resolve the
formation of galaxy disks over the same mass scales as in our sample
\citep[e.g.][]{Vogelsberger-14,Schaye-15, Pillepich-18}. These models follow
the cooling and condensation of gas in galaxy halos. The formation of molecular
gas is often incorporated as post-processing step \citep[e.g.][]{Diemer-19a}.
Comparisons with H{\,\sc i} scaling relations such as those presented in this
section then constrain gas accretion and condensation processes in the models,
which depend sensitively on processes such as supernova and AGN feedback.

\section{Conclusions}

Using the rest-frame near-ultraviolet (NUV)-optical  colour, $NUV-r$, as a
proxy for the H{\,\sc i}-to-stellar mass ratio,  $M_{\rm HI}/M_\ast$, we have
estimated the H{\,\sc i} gas mass  for a sample of $\sim7,000$ galaxies in the
DEEP2 survey with redshifts in the range $0.75<z<1.4$ and stellar masses above
$\sim10^{10}$\,$\rm M_\odot$. The correlation between $M_{\rm HI}/M_\ast$ and $NUV-r$ is
calibrated with a sample of 660 galaxies in the local Universe that have
optical and $NUV$ photometry from SDSS and GALEX, as well as 21-cm emission data
from HyperLeda. With these H{\,\sc i} mass estimates, we have calculated H{\,\sc
i} mass functions (HIMFs) and cosmic H{\,\sc i} gas densities ($\Omega_{\rm HI}$),
and we have investigated the correlations between the specific star formation rate
(SFR/$M_\ast$), the H{\,\sc i}-to-stellar mass ratio ($M_{\rm HI}/M_\ast$) and the
H{\,\sc i} gas depletion time ($t_{dep}({\rm HI})\equiv M_{\rm HI}/{\rm SFR}$).

Our main conclusions can be summarized as follows.
\begin{itemize}

  \item For galaxies with stellar masses above $\sim10^{10}$\,$\rm M_\odot$, 
    the  HIMF at $z\sim1$ is quite similar to
    the local HIMF, while the low-mass end slope cannot be constrained due to the
    lack of low stellar mass galaxies in our sample.

  \item Integrals of our HIMFs suggest the lower limit of the H{\,\sc i} mass
    density at $z\sim1$ to be $\Omega_{\rm HI, limit}\sim2.1\times10^{-4}\,
    h_{70}^{-1}$, which is significantly smaller than the estimates obtained by
    other authors based on DLAs at the same redshift. However, our work confirms
    the result of \cite{Kanekar-16}, that massive star-forming galaxies do not
    dominate the neutral gas at $z\sim1$. 

  \item Galaxies with $M_*\sim 10^{11}$\,$\rm  M_\odot$ at $z=1$ have large H{\,\sc
    i} gas fractions, which are about 3$-$4 times higher than local galaxies,
    indicating the neutral gas fraction evolves from  $z=1$ to the present
    epoch. It is found that the slope of the correlation between 
    H{\,\sc i} gas fraction and stellar mass also has a modest evolution with redshift.

  \item The quantity $M_{\rm HI} / {\rm SFR}$ exhibits very large scatter,
         and the scatter increases from a factor of 5--7 at $z=0$ to factors in
         close to a hundred at $z=1$. This implies that there is no relation between
         H{\,\sc i} gas and star formation in high redshift galaxies. The 
         H{\,\sc i} gas must be linked to cosmological gas accretion processes.

\end{itemize}

As discussed previously, all the results obtained in this work rely on the
assumption that the same scaling relation between H{\,\sc i} mass fraction and
$NUV$-optical colour applies at higher redshifts. Although our measurements of
gas fraction and cosmic H{\,\sc i} mass mass density are consistent with that
derived from co-adding method or Mg{\,\sc ii} absorbers, it would be very
useful to obtain small samples of galaxies with H{\,\sc i} measured at $z=1$ to
validate that the scaling relations hold at this redshift. This
goal should be achieved with the help of the future radio telescopes such as
the next-generation VLA (ngVLA) and the Square Kilometre Array (SKA). From the
linear fitting relations between $M_{HI}/M_*$ and $M_*$ for the galaxies in two
high-z bins in Fig.~\ref{Fig:Mass_etal}, a lower limit of $M_{HI} \sim 2.7
\times 10^{10} M_\odot$ is required to derive the scaling relations covering a
wide range of $M_*$ from $10^9$ to $10^{11}\,M_\odot$ at $z\sim1$. Using the
equation  $\frac{M_{HI}}{\rm M_\odot} = 2.356 \times 10^5 D_{\rm Mpc}^2 \times
S_{21}$, where $D_{\rm Mpc}$ is the distance to the galaxy in Mpc, the
integrated flux density detection threshold $S_{\rm 21,th}$ can be calculated as
$2.6 \times 10^{-3}\,\rm Jy\, km\,s^{-1}$. Taking this threshold, sufficient
number of galaxies can be detected at $z\sim1$ according to the simulations of
SKA detection rates \citep{Obreschkow-09b}.

\section*{Acknowledgement}

W.Z. is grateful to MPA for hospitality when this work was being completed.
W.Z. deeply appreciate that Cheng Li has given a lot of guidance and
recommendations for this work.  W.Z. thanks Kevin Bundy for kindly sharing the
catalogue of $K_{\rm s}$ magnitude and stellar mass. W.Z. thanks Zheng Zheng,
and Qi Guo for helpful suggestions. This work is supported by the Joint
Research Fund in Astronomy (No. U1531118) under cooperative agreement between
the National Natural Science Foundation of China (NSFC) and Chinese Academy of
Sciences (CAS). This work is also sponsored by NSFC (No. 12090041, 12090040,
10903011, 11173045, 11733006), the National Key R\&D Program of China grant No.
2017YFA0402704, and the Guangxi Natural Science Foundation (No.
2019GXNSFFA245008). J.F. acknowledges the support by the Youth innovation
Promotion Association CAS and Shanghai Committee of Science and Technology
grant No.19ZR1466700 and the support by the Joint Research Fund in Astronomy
(No. U1531123). This work has made use of data from the SDSS and SDSS-II, and
the HyperLeda database. 


%
\bibliographystyle{aa} 
\bibliography{ref}


\label{lastpage}

\end{document}